\documentclass[12pt,a4paper]{iopart}

\usepackage{iopams,braket}
\usepackage{graphicx,overpic,floatrow}
\usepackage{amsmath,bm}
\usepackage[caption=false]{subfig}
\usepackage{color, colortbl}
\definecolor{Gray}{gray}{0.9}

\usepackage[colorlinks=true,citecolor=blue,linkcolor=black]{hyperref}

\usepackage{cite}
\usepackage{physics}
\usepackage{ascmac}
\usepackage{dcolumn}
\usepackage{here}
\usepackage{wrapfig}
\usepackage[dvipsnames]{xcolor}

\begin{document}

\title{Condensate-mediated dimerization of impurities in atomic BECs
}

\author{Hoshu Hiyane}
\address{Quantum Systems Unit, Okinawa Institute of Science and Technology Graduate University, Onna, Okinawa 904-0495, Japan}
\ead{hoshu.hiyane@oist.jp}

\author{Thom\'as Fogarty}
\address{Quantum Systems Unit, Okinawa Institute of Science and Technology Graduate University, Onna, Okinawa 904-0495, Japan}

\author{Jose Carlos Pelayo}
\address{Quantum Systems Unit, Okinawa Institute of Science and Technology Graduate University, Onna, Okinawa 904-0495, Japan}

\author{Thomas Busch}
\address{Quantum Systems Unit, Okinawa Institute of Science and Technology Graduate University, Onna, Okinawa 904-0495, Japan}

\vspace{10pt}
\begin{indented}
\item[] \today
\end{indented}

\begin{abstract}
    We show that strongly correlated impurities confined in an optical lattice can form localized, molecule-like dimer states in the presence of a Bose–Einstein condensate (BEC).
    By systematically studying the effect of the lattice potential on this mixture, we reveal the two roles of the condensate in assisting the formation of dimerized impurities: mediating the attractive interaction among impurities and rescaling the lattice potential of impurities.
    At strong coupling between the impurities and the condensate, the two mechanisms cooperate to induce a structural transition, resulting in the rearrangement of dimers.
    We also show that the nonequilibrium dynamics of these states can be interpreted as a dimerized soliton train.
\end{abstract}

\vspace{2pc}

\section{Introduction}

Low-dimensional systems often show emergent quantum phenomena that can be contrasted with the behavior of their three-dimensional counterparts, requiring a fundamentally different theoretical approach~\cite{Giamarchi_2003,pitaevskii_stringari2016,Mistakidis2023Nov}. 
A prime example is the strongly interacting Tonks–Girardeau (TG) regime, where a one-dimensional Bose gas exhibits fermion-like properties despite obeying bosonic statistics~\cite{Girardeau:60}. 
As the spatial density profile of the TG gas mirrors that of a non-interacting Fermi gas with the same particle number, the Fermi wavevector emerges as a characteristic length scale. 
When an optical lattice is applied with a wavevector commensurate with the TG gas density, the system can undergo a quantum phase transition from a delocalized, superfluid-like (SF) state to a localized insulating state, even for vanishingly small lattice depths~\cite{Buchler_2003}. This phenomenon, known as the pinning transition, arises from the commensurate lattice effectively “pinning” the TG atoms in place~\cite{haller_2010}.

In contrast, for a weakly interacting one-dimensional Bose gas at sufficiently low temperature, a (quasi-) Bose--Einstein condensate (BEC) emerges, which can be described by the celebrated Gross--Pitaevskii equation (GPE)~\cite{Book:Popov_1983,Mora_2003,pitaevskii_stringari2016}.
For attractively interacting one-dimensional Bose-condensed systems, the mean-field GPE admits localized solutions in the form of bright solitons as a ground state, as well as highly excited soliton-train states~\cite{McGuire1964May,Carr_2000_2,Kanamoto_2005}.
When such an attractively interacting condensate is subjected to an external potential having a geometric length scale, e.g.\ a double-well potential or a lattice potential, the solution that respects the symmetry of the system is known to be unstable beyond a certain self-attraction strength, resulting in a spontaneous localization in one of the wells.
Such spontaneous breaking of the symmetry has been studied in the context of a bosonic Josephson junctions~\cite{Smerzi_1997,Raghavan_1999,Mazzarella_2010,wysocki_2024}, which are a manifestation of macroscopic quantum phenomenon emerging due to the coherent matter-wave nature of the Bose-condensed gas~\cite{Josephson1962Jul,Book:Mahan2000,pitaevskii_stringari2016}.

Interestingly, immersing a TG or a noninteracting Fermi gas into a BEC, allows one to engineer the mean-field potential to create a net attractive interaction for the immersed species, which can also lead to a spatially self-localized state~\cite{Karpiuk_2004,Salerno_2005,Karpiuk_2006,Santhanam_2006,DeSalvo_2017,Rakshit_2019_NJP,Rakshit_2019_scipost,Desalvo_2019,Keller_2022,hiyane_2024}.
Here, by applying an optical lattice, we explore the interplay between the commensurate pinning effect and self-localization of TG or fermionic impurities immersed in a BEC.
First, we demonstrate that the BEC-mediated self-attraction leads to the formation of a molecule-like dimer state in the ground state of the impurities by applying the optical lattice {\it selectively} to the impurities species.
Based on this result, we investigate the ground state of the BEC-TG mixture when {\it both} components are subject to the same optical lattice, and find that a structural transition occurs for the dimerized impurities as a function of the coupling strength.
Finally, we study the nonequilibrium properties of dimerized impurities by examining the dynamics following the sudden removal of the optical lattice.
We show that at weak coupling, the system thermalizes within the BEC bath, while at strong coupling, when dimers are tightly localized, the system exhibits quasi-integrable behavior and can be thought of as a dimerized soliton-train.

This manuscript is organized as follows.
We first introduce the governing equations and the effective model in Section~\ref{sec:model}.
In Section~\ref{sec:groundstate_Voptonlyinimpurities}, the BEC-impurity in a species-selective optical lattice potential is studied, while in Section~\ref{sec:groundstate_Voptinboth} we investigate the situation where both components experience the lattice. 
Finally, we discuss the dynamics of the dimerized impurities in Section~\ref{sec:quench_dynamics} and conclude in Section~\ref{sec:conclusion}. Additional discussions on the relation of our system to a Bose–Josephson junction are provided in \ref{appendix:double_well}.

\section{Modeling TG--BEC mixture trapped by optical lattices\label{sec:model}}

The system we study consists of two coupled one-dimensional quantum gases: a weakly interacting Bose–Einstein condensate and a small system of strongly correlated impurities in the Tonks–Girardeau regime. 
The BEC is described by the mean-field Gross–Pitaevskii equation for the condensate wavefunction $\Psi(x,t)$ \cite{pitaevskii_stringari2016}, while the TG impurities can be mapped to a system of non-interacting spinless fermions \cite{Girardeau:60}, characterized by single particle states $\phi_n(x,t)$. 
Neglecting the correlations between impurities and BEC, we model the interaction of the two components as a repulsive density-density coupling with strength $\gamma$, scaled relative to the mean-field interaction strength of the condensate $g$~\cite{Keller_2022,hiyane_2024,Zschetzsche_2024}. Assuming equal masses $m$ for both components, the coupled equations governing the system are given by
\begin{align}
    i\pdv{t}\Psi(x,t)=&\left[-\frac{1}{2}\pdv[2]{x}+V_{\rm BEC}(x)+|\Psi(x,t)|^2+\gamma\rho(x,t)\right]
    \Psi(x,t),
    \label{eq:GPE}\\
    i\pdv{t}\phi_n(x,t)=&\bqty{-\frac{1}{2}\pdv[2]{x}+V_{\rm imp}(x)
    +\gamma|\Psi(x,t)|^2}\phi_n(x,t),
    \label{eq:SPSE}
\end{align}
where $V_{\rm BEC}(x)$ and $V_{\rm imp}(x)$ denote the external potentials for the BEC and the impurities, respectively, and $\rho(x,t)$ is the density of the TG impurities.  
The equations and all the quantities introduced here and hereafter are scaled to a dimensionless form in terms of the characteristic length scale $x_0 = \hbar^2 / (mg)$, which is associated with the condensate healing length, and the characteristic time scale $t_0 = mg^2 / \hbar^3$.

The many-body ground state wavefunction of the Tonks--Girardeau gas at zero temperature can be obtained using the Bose--Fermi mapping $\Phi(\bm{x},t) = |\Phi_{\rm F}(\bm{x},t)|$, where $\Phi_{\rm F}(\bm{x},t) = \det[\phi_n(x_m,t)]_{1 \leq n,m \leq N}/\sqrt{N!}$ represents the many-body wavefunction of the mapped Fermi gas, and $\bm{x} = (x_1, \cdots, x_N)$ with $N$ being the number of TG impurities~\cite{Girardeau:60}.  
The total TG density is then expressed as $\rho(x,t) = \sum_{n=1}^{N} |\phi_n(x,t)|^2$. Assuming stationary solutions of the form $\phi_n(x,t) = \phi_n(x)e^{-iE_nt}$ and $\Psi(x,t) = \Psi(x)e^{-i\mu t}$, where $E_n$ and $\mu$ denote the single-particle energies and chemical potential, respectively, equations~\eqref{eq:GPE} and~\eqref{eq:SPSE} can be solved self-consistently to determine the ground state of the coupled system. 
To study the properties of a few TG bosons immersed in a bulk of a condensate, the calculations are performed with closed boundary conditions for the BEC and open boundary conditions for the impurities within a system size of length $L = 50$, and the number of condensate particles and impurities are chosen as $N_{\rm BEC} = 10^4$ and $N \sim 10$ such that $N_{\rm BEC} \gg N$.

In this high density-imbalance regime, the Thomas–Fermi approximation provides a suitable framework to describe both the ground state and the dynamics of the impurities~\cite{hiyane_2024}. 
It allows for the construction of an effective model for the impurities alone~\cite{Keller_2022, hiyane_2024}, so that the solution to equation~\eqref{eq:GPE} can be written as
\begin{align}
	|\Psi(x,t)|^2\approx\mu_{\rm TF}-\gamma\rho(x,t)-V_{\rm BEC}(x),
	\label{TFA}
\end{align}
where the equilibrium chemical potential within the TFA is given by
\begin{align}
	\mu_{\rm TF}
    =\frac{N_{\rm BEC}}{L}+\frac{\gamma N}{L}+\int \frac{dx}{L}V_{\rm BEC}(x).
\end{align}
Using this framework, the Schr\"odinger equation for the impurities, equation~\eqref{eq:SPSE}, can be reformulated as
\begin{align}
    i\pdv{t}\phi_n(x,t)
    =\bqty{-\frac{1}{2}\pdv[2]{x}
    +V_{\rm imp}-\gamma V_{\rm BEC}
    -\gamma^2\rho
    + \gamma\mu_{\rm TF}
    }\phi_n(x,t).
    \label{eq:TFASE}
\end{align}

This expression represents a system of $N$-coupled nonlinear Schrödinger equations ($N$-CNLSE), which describe the ground state and dynamical properties of the impurities, when excitations in the BEC can be neglected~\cite{hiyane_2024}. 
The influence of the condensate is twofold: it mediates interaction among impurities through the term $-\gamma^2 \rho$ and it modifies the single-particle trapping potential seen by the impurities through the  effective trapping potential
\begin{align}
    V_{\rm eff}=V_{\rm imp}(x)-\gamma V_{\rm BEC}(x).
    \label{eq:effective_lattice_trap}
\end{align}


\section{\label{sec:groundstate_Voptonlyinimpurities}Impurities in a species selective lattice}

As can be seen from equation~\eqref{eq:TFASE}, the ground state properties of the impurities depend on the external potential $V_{\rm imp}$, the rescaled trapping potential of the BEC $-\gamma V_{\rm BEC}$, and BEC-mediated self-interaction among impurities $-\gamma\rho$.
The latter is of attractive character and therefore able to induce localization effects among impurities~\cite{Keller_2022,hiyane_2024}.

As the single particle potential, we consider an optical lattice trap described by
\begin{align}
    V_{\rm opt}(x)=V_0\sin^2(k_{\rm r}x+\theta),
    \label{eq:Vopt}
\end{align}
where $k_{\rm r} = N_{\rm s}\pi / L$, with $N_{\rm s}$ representing the number of lattice sites that is superposed on a hard-wall box with length $L$. The lattice depth $V_0$ is scaled in terms of the recoil energy $E_{\rm r}$, defined as $E_{\rm r} = k_{\rm r}^2 / 2$. The phase of the optical lattice, $\theta$, is chosen such that $\theta = 0$ for an odd number of lattice sites and $\theta = \pi/2$ for an even number, ensuring the lattice contains an integer number of sites within the system.

\begin{figure}[tb]
    \centering
    \includegraphics[width=\linewidth]{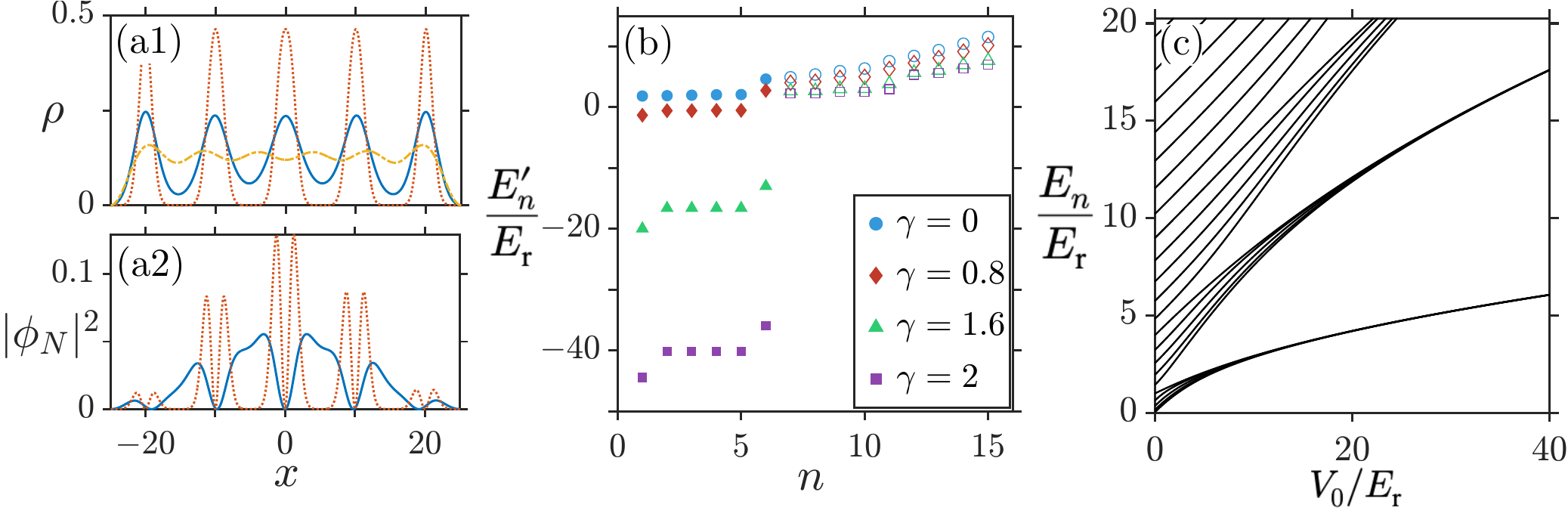}
    \caption{
    (a1) Ground state density profile of a TG gas with $N=6$ particles in a lattice potential with $N_{\rm s}=5$ sites and (a2) the highest occupied single particle states for a shallow lattice $V_0=5E_{\rm r}$ (solid line) and for a deep lattice $V_0=52.3E{\rm r}$  (dotted line).
    The density without a lattice potential is shown as the dot-dashed line for reference.
    (b) Rescaled single particle energies of TG particles $E'_n=E_n-\gamma\mu_{\rm TF}$ (see equation~\eqref{eq:TFASE}) in the presence of an optical lattice with depth $V_0=5E_{\rm r}$. 
    (c) Single particle energies of the TG bosons ($\gamma=0$) as the function of the lattice depth.
    }
    \label{fig:pinningtr}
\end{figure}

We begin our discussion by briefly recalling the physics that governs the impurity system alone, which is described by $\gamma=0$ in equation~\eqref{eq:SPSE}~\cite{Buchler_2003,haller_2010,Mikkelsen_2018}. The potential is then $V_{\rm imp}=V_{\rm opt}$ with a hard-wall box boundary conditions and for concreteness we consider an incommensurate setup with $N_{\rm s} = 5$ lattice sites and $N = 6$ TG gas particles at zero temperature.
The densities for increasing lattice strengths are shown in Fig.~\ref{fig:pinningtr}(a1), where one can see that the density is delocalised 
in the absence of the lattice, but localises into a state having $N_{\rm s}$ maxima once the lattice is switched on. However, as the density between different lattice sites possesses a finite overlap for $V_0=5E_{\rm r}$, this state can be interpreted as being constructed by $N_{\rm s}$ low-lying states that are tightly pinned by the optical lattice and that contribute the $N_{\rm s}$ maxima in the density profile, while the highest occupied state $\phi_N$ is still delocalized over the box.
The distribution of this state $|\phi_N|^2$ is shown in panel (a2) for the two lattice depths used in panel (a1). 
One can see that (i) the discrete translational symmetry that the optical lattice~\eqref{eq:Vopt} provides is explicitly broken due to the external hard-wall box trap, resulting in the overall maximum of $|\phi_N|^2$ at the trap center,
(ii) for deeper lattices the state $\phi_N$ is more localized in each minimum of the lattice sites,
and
(iii) for the shallow lattices, this state is more delocalized with contributions from the regions between the lattice sites, allowing them to avoid overlapping with the particles that are localized around the lattice sites' minima.
Correspondingly, the lowest $N_{\rm s}$ single particle energies of the TG atoms are nearly degenerated forming an energy band that is well-separated from the $n=N_{\rm s}+1$ state, leading to a gap in the spectrum (see panel (b)).
This energy gap between $n=N_{\rm s}$ and $n=N_{\rm s}+1$ becomes larger for deeper lattice depths, and higher bands start to appear for sufficiently deep lattice depths, as can be seen from panel (c).

\begin{figure}[tb]
    \includegraphics[width=\linewidth]{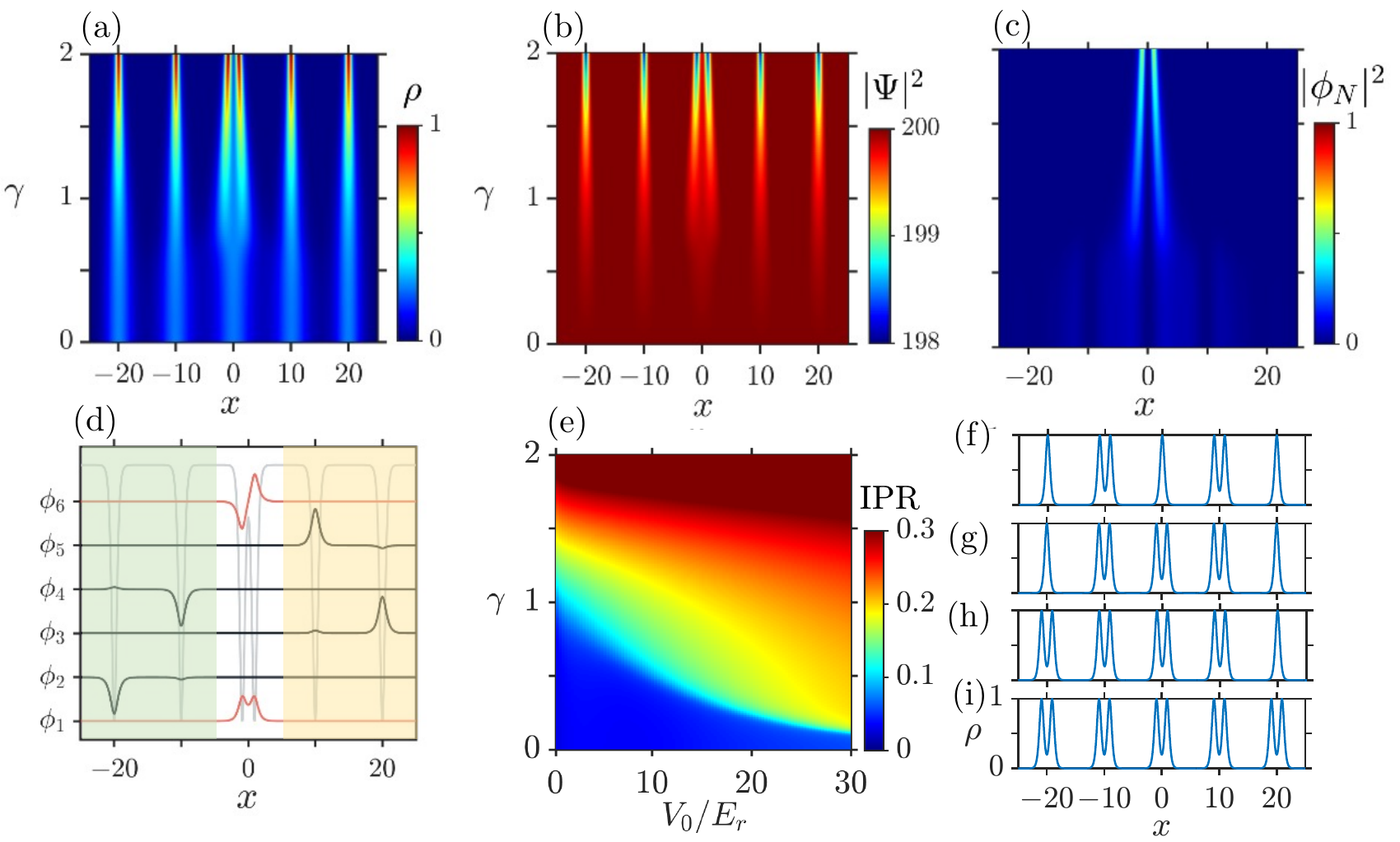}
    \caption{
    The ground state density of the (a) impurities and (b) BEC, and the (c) highest occupied single particle states $|\phi_N|^2$ are shown as a function of position $x$ and inter-species interaction strength $\gamma$.
    (d) Single-particle states at $\gamma = 2$, including the schematic effective potential formed by the BEC in the background. 
    (e) The IPR of the $\phi_N$ state as a function of the $\gamma$ and the lattice depth $V_0$.
    (f–i) The density profiles of the dimer state are displayed for (f) $N = 7$, (g) $N = 8$, (h) $N = 9$, and (i) $N = 10$, respectively. 
    The system consists of $N_{\rm BEC} = 10^4$ condensate atoms with a periodic boundary conditions, while the TG-impurities are in a hard-wall box with a length of $L = 50$.
    The impurities experience a species-selective lattice trap, described by equation~\eqref{eq:Vopt}, with a lattice depth of $V_0 = 5E_{\rm r}$. 
    The rest of the parameter choices are the same as in Fig.~\ref{fig:pinningtr}.
    } 
    \label{fig:speciesselective_density}
\end{figure}

To study the effect of the condensate onto this pinning state, we now turn on the coupling strength $\gamma$ and first focus on the case where the BEC atoms do not experience the optical lattice potential, i.e\ $V_{\rm BEC} = 0$ and $V_{\rm imp} = V_{\rm opt}$.
The ground state density profile of impurities is shown in Fig.~\ref{fig:speciesselective_density} (a) for $V_0 = 5E_{\rm r}$.
One can see that, as the inter-species interaction strength $\gamma$ increases, the impurities already pinned in the optical lattice become more localized due to the repulsive mean-field potential generated by the BEC.
This can also be seen from the appearance of $N_{\rm s}$-minima in the condensate density for $\gamma\gtrsim 0$ in Fig.~\ref{fig:speciesselective_density} (b), which mirrors the density distribution of the impurities.

Moreover, the impurity density starts localizing towards the trap center with increasing $\gamma$, and for a coupling strength larger than $\gamma\gtrsim 0.8$, a double peaked maximum emerges at the center of the trap.
From the single particle states in the strong coupling limit $\gamma=2$ shown in panel (d), we can see that the states $\phi_1$ and $\phi_N$ are both localised in the central lattice site around $x=0$, and in panel (c) we show how the initially delocalized state $|\phi_N|^2$ starts to localize towards the box center. 
This localization occurs because the BEC-mediated attraction among the impurities (described by $N$-CNLSE~\eqref{eq:TFASE}) is stronger around the center of the trap due to the overall maximum in the density caused by the box trap (see Fig.~\ref{fig:pinningtr}(a), particularly (a2)), and in the strong coupling limit, $\phi_N$ prefers to localize around $x = 0$. 
A similar transition occurs for the state $\phi_1$, except in this case there is no node at the trap center as it is the lowest energy state.
On the other hand, the higher energy state $\phi_N$ possesses a node at $x=0$ resulting in a profile presented in panel (d).

We emphasize that such dimer states cannot exist without a BEC, even for the particles in deep lattice wells.
For the reference, the $|\phi_N|^2$ presented in Fig.~\ref{fig:pinningtr}(a2) is confined in the deep lattice whose depth is taken to be comparable to the BEC matter-wave lattice estimated for Fig.~\ref{fig:speciesselective_density}(b) at $\gamma=2$, that is $\gamma(\max|\Psi|^2-\min|\Psi|^2)\approx 52.3E_{\rm r}$.
Instead of the definitive double occupation at the single site, the $\phi_N$ state localizes at each site with a visible finite size effect, signaling the absence of the bound dimer states.

Thus, through the mean-field coupling which appears as an effective single particle potential in equation~\eqref{eq:SPSE}, the condensate binds two TG atoms resulting in the formation of the dimer state.
This can be seen from the ``bare'' single particle energy of impurities which can be obtained by removing the large background energy of the condensate, $E'_{n}=E_n - \gamma\mu_{\rm TF}$ (see equation~\eqref{eq:TFASE}), in Fig.~\ref{fig:pinningtr}(b).
Before the formation of the dimer $(\gamma\lesssim0.8)$, the change in the rescaled energy $E'_n$ is almost negligible compared to after the formation occurs 
($\gamma > 0.8$); Once the dimer is formed, the energy of the occupied state $E'_{1\leq n\leq N}$ significantly changes and starts to drop, signaling the emergence of a bound state induced by the BEC matter-wave potential.
In particular, the energies $E'_1$ and $E'_{N_{\rm s}+1}$ increasingly deviate from the first band ($E’_{1 \leq n \leq N_{\rm s}}$) and second band ($E'_{N_{\rm s}+1 \leq n \leq 2N_{\rm s}}$), respectively, as the dimer state starts forming.

The states $\phi_{2 \leq n \leq N_{\rm s}}$, on the other hand, remain nearly degenerate for all values of $\gamma$, corresponding to tightly localized wavefunctions that form the four density maxima without double occupation shown in Fig.~\ref{fig:speciesselective_density}(a) and (d).
Due to the mediated self-attraction, each impurity localizes at one of the lattice minima except at the central site, which the dimer occupies 
(for further discussion of Fig.~\ref{fig:speciesselective_density}(d), see \ref{appendix:double_well}).
Such sharp localization manifests in a vanishing overlap with the adjacent site, which allows one to treat each of the four particles approximately as a collection of single particles and reduces $\rho\to|\phi_n|^2$ in $N$-CNLSE~\eqref{eq:TFASE}.
Moreover, when the coupling energy dominates over underlying lattice energy in the strong coupling limit, one can set $V_{\rm eff}\approx0$ in the $N$-CNLSE~\eqref{eq:TFASE} and \eqref{eq:effective_lattice_trap}.
This means that the $N$-CNLSE~\eqref{eq:TFASE} reduces to $N$ free NLSEs with attractive self-interaction, which possesses the well-known bright soliton solutions with energy $E'_{2\leq n\leq 4}=-\gamma^4/8$~\cite{Keller_2022,Carr_2000_2}, that is $E'_{2\leq n\leq 4} \approx -16E_{\rm r}$ and $-40E_{\rm r}$ for $\gamma=1.6$ and 2, respectively, which agree quite well with Fig.~\ref{fig:speciesselective_density}(b).

Let us now investigate the relationship between lattice depth and localization, for which we compute the inverse participation ratio (IPR) of the highest occupied state $\phi_N$, defined as ${\rm IPR} = \int dx |\phi_N(x)|^4 dx$, and shown in Fig.~\ref{fig:speciesselective_density}(e). 
The IPR quantifies the localization of the state $\phi_N$ as a function of the coupling strength $\gamma$ and the optical lattice depth $V_0/E_{\rm r}$. One can see that the IPR increases monotonically for all values of $V_0$,
which indicates the formation of localized dimer states.
This formation occurs earlier for deeper lattice depths, as the BEC-mediated self-attraction becomes stronger there.

We next discuss the arrangement of dimers as a function of the number of impurities. Figures~\ref{fig:speciesselective_density}(f)–(i) display the density profiles of the impurities for $N = 7$ to $N = 10(=2N_{\rm s})$ at $\gamma = 2$.
The dimers are composed of strongly interacting TG bosons; thus, they prefer to minimize the overlap between them and also with the edge of the box, for $N=7$ they form in next neighbouring sites and away from the edge. Adding one more particle, $N=8$, then requires to fill an additional site, which is the one at the center of the box in order to avoid the edge while maintaining symmetry.

For $N = 9$, the self-attraction among impurities makes the symmetric arrangement of dimers unstable. Instead, the system spontaneously breaks symmetry, forming a dimer at either the leftmost or rightmost well. To control this symmetry breaking, we introduce a weak linear potential, $V_{\rm L} = 10^{-10}x$, which is orders of magnitude smaller than any other energy scale in the system. This small perturbation directs the dimer formation to the leftmost well (for a detailed discussion on symmetry breaking, see \ref{appendix:double_well}).

Finally, for $N = 10(=2N_{\rm s})$, the impurities completely fill the lowest two bands, resulting in a doubly occupied state throughout all the $\gamma$.
However, even in this case, it is worth noting that the effect of the coupling with BEC is not limited to inducing further localization but also binds the two TG atoms, leading to emerging properties such as the formation of solitonic dynamics as we will discuss in Section~\ref{sec:quench_dynamics}.

Before closing this section, let us comment on the relation of this impurity-BEC mixture to a bosonic Josephson junction system. 
The observed self-localization has a strong connection to the situation of an attractive condensate in a bosonic-Josephson junction.
This is particularly clear for the single impurity case; the effective governing equation of the impurity is simply given by NLSE with self-attraction ($n=N=1$ in $N$-CNLSE~\eqref{eq:TFASE} and $\rho=|\phi_1|^2$).
The underlying lattice structure can be viewed as the junctions where the impurity hops from one site to another while experiencing the self-attraction $-\gamma^2|\phi_1|^2$.
Hence, the system in this case is effectively the same as the attractively interacting BEC confined in a bosonic Josephson junction~\cite{Smerzi_1997,Raghavan_1999,Mazzarella_2010,wysocki_2024}.
Then, as the $N$-CNLSE extends the NLSE, the mechanism of the localization in $N$-impurity system can also be viewed in terms of this effective picture.
In \ref{appendix:double_well}, we push this analogy further and discuss more details about the emergence of dimers and spontaneous symmetry breaking, and in the following, we further discuss the properties of the impurities in a condensate.


\section{
BEC--TG mixture in the optical lattice\label{sec:groundstate_Voptinboth}}

\begin{figure}[ht]
    \centering
    \includegraphics[width=\linewidth]{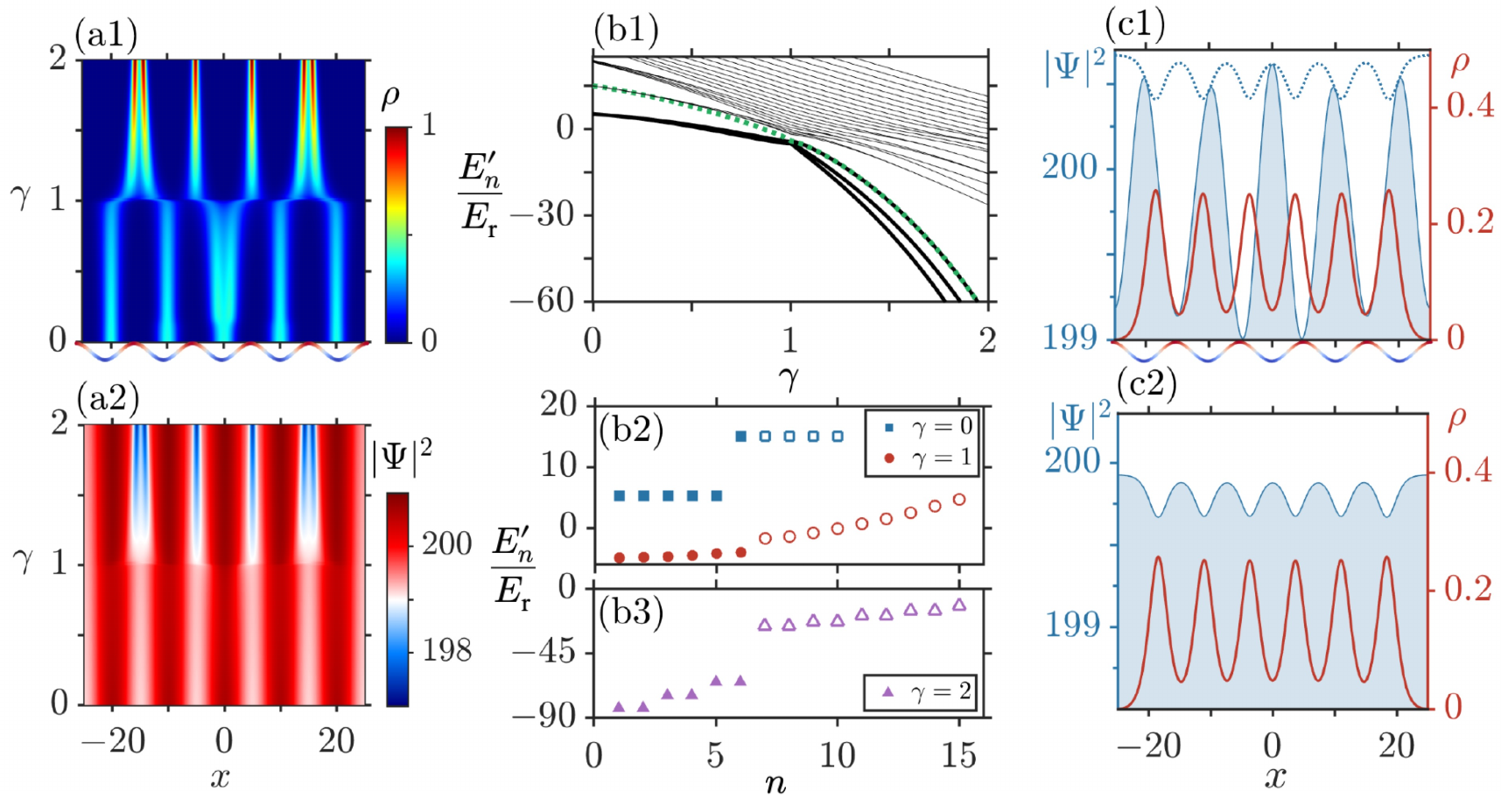}
    \caption{
    $N=6$ impurities in a condensate trapped by optical lattice.
    (a) Density of (a1) impurities and (a2) BEC as a function of $\gamma$.
    (b1) Lowest 30 rescaled single particle energies $E'_{n}$ as a function of $\gamma$. 
    The thick and thin black lines correspond to the states $1\leq n<N$ and $n>N$ respectively, while the green dotted line corresponds to the state $n=N$.
    (b2) and (b3) are the lowest few single particle energies for the specific coupling strength $\gamma$ with occupied states indicated by the filled mark.
    (c1) Density of BEC $|\Psi|^2$ (blue filled curve and left axis) and impurities (red line and right axis) for $\gamma=1$.
    The blue dotted line is the effective potential for the impurities $V_{\rm imp} + \gamma |\Psi(x)|^2$ (see equation ~\eqref{eq:SPSE}).
    (c2) Corresponding to (c1), we show the density of the BEC and impurities for $\gamma=1$ and $V_0=0$.
    From (a) to (c1), the lattice parameter choices are $V_{\rm imp}=V_{\rm BEC}=V_{\rm opt}$ with $V_0=30E_{\rm r}$ and $N_{\rm s}=5$, which are schematically shown below the frame of (a1) and (c1). 
    The rest of the parameters are the same as Fig.~\ref{fig:speciesselective_density}.
    }
    \label{fig:VoptinbothGS}
\end{figure}

In the previous section, we demonstrated that species-selectively trapped impurities immersed in a BEC can form molecule-like dimer states due to the BEC-mediated attraction among impurities, as described by the $N$-CNLSE in equation~\eqref{eq:TFASE}. Based on these findings, we now examine the TG-BEC mixture in an optical lattice where $V_{\rm imp} = V_{\rm BEC} = V_{\rm opt}$, in equations~\eqref{eq:GPE} and \eqref{eq:SPSE}.
In this scenario, the impurities experience an effective trap defined by equation~\eqref{eq:effective_lattice_trap} as  $V_{\rm eff}=(1-\gamma)V_{\rm opt}$, and the role of the condensate becomes twofold: it not only provides effective self-attraction among the impurities but also rescales the effective trapping potential of the impurities. Specifically, the nature of the effective trap depends on the coupling strength $\gamma$ and for weak coupling $(\gamma < 1)$, the effective optical lattice is repulsive, whereas for strong coupling $(\gamma > 1)$, it becomes attractive.

The ground state density of the impurities and the BEC is shown in Figs.~\ref{fig:VoptinbothGS} (a1) and (a2) for a relatively deep lattice with $V_0 = 30E_{\rm r}$ and $N_{\rm s} = 5$. Initially, at $\gamma \approx 0$, the density profiles of both the impurities and the BEC exhibit five maxima located around the minima of the external optical lattice (blue colored region in the schematic lattice below panels (a1) and (c1)).

In the weak coupling regime $(\gamma \ll 1)$, the influence of the trapping potential on the BEC is small since it appears as $\gamma V_{\rm opt}$ in the effective potential of impurities $V_{\rm eff}$~\eqref{eq:effective_lattice_trap}, and the ground state configuration of the impurities is primarily determined by the interplay between the lattice trap $V_{\rm imp}$ and the BEC-mediated self-attraction. 
This limit is then similar to the situation of the species-selective lattice trap for the impurities discussed in the previous section.
Therefore, a dimer is formed around the trap center when the coupling becomes sufficiently strong, that is $\gamma\approx 0.2$ for the lattice depth considered here ($V_0=30E_{\rm r}$), consistent with the phase diagram in Fig.~\ref{fig:speciesselective_density}(e) obtained for the species-selective trap setup.

As the coupling strength increases to $\gamma \approx 1$, the central dimer splits, leading to the formation of new dimers with different arrangements in the strong coupling limit $(\gamma \gg 1)$. At this stage, the effective potential, given by equation~\eqref{eq:effective_lattice_trap}, becomes an attractive optical lattice with four lattice sites and two half-lattice sites near the edges of the box (red colored region in the schematic lattice below panels (a1) and (c1)).
As discussed in Figs.~\ref{fig:speciesselective_density}(f) and (g), since the systems aims to minimize the overlap between adjacent lattice sites while avoiding the box edges, the two dimers appear symmetrically. 

This structural transition is accompanied by a dramatic change in the energy bands observed in the single-particle spectrum. In Fig.~\ref{fig:VoptinbothGS}(b1), we plot a few of the lowest rescaled single-particle energies $E’_n$ as a function of the coupling strength $\gamma$.
At $\gamma = 0$, the five lowest energy states are nearly degenerate due to the deep optical lattice, forming well-defined bands with clear band gaps. This is also evident in Fig.~\ref{fig:VoptinbothGS}(b2), which depicts the spectrum at $\gamma = 0$.
In the strong coupling limit $(\gamma = 2)$, the BEC-mediated attractive lattice reorganizes the impurities, resulting in the formation of new energy bands distinct from those observed at $\gamma = 0$. As shown in panel (b3), each band in this regime consists of two nearly degenerate states, consistent with the density profile observations.
Notably, panel (b1) illustrates that as the coupling constant $\gamma$ increases, the initially large band gaps present at $\gamma = 0$ begin to close. At $\gamma = 1$, these gaps have fully closed, and a new band gap emerges between $E_N$ and $E_{N+1}$, as also shown in panel (b2).
This behavior arises because the effective trap potential for the impurities, described by equation~\eqref{eq:effective_lattice_trap}, is exactly canceled at $\gamma = 1$, and the resulting $N$-CNLSE contains no external single-particle potential. Consequently, a self-pinning state emerges, driven by the competition between the Pauli exclusion principle and the BEC-mediated effective attraction~\cite{Keller_2022}, which will be discussed soon below.
A distinctive feature of impurities in the self-pinning state is the absence of multiple band gaps, irrespective of the depth of the matter-wave lattice formed by the BEC~\cite{Keller_2022}.

This observation motivates a comparison of the density profiles of the composite system at the transition point $\gamma = 1$ for the cases $V_0 = 0$ and finite $V_0$. The density profile of the self-pinning state $(V_0 = 0)$ at $\gamma = 1$ is shown in panel (c2).
Due to the repulsive mean-field density-density coupling, the density profile of the BEC (blue-filled curve) is modulated, acquiring a periodicity determined by the Fermi wavevector of the impurities. This modulation acts as a matter-wave lattice that commensurately traps the impurities, leading to the formation of band gaps in the single-particle energy spectrum at $n = N$.
Figure~\ref{fig:VoptinbothGS} (c1) shows the density profiles of the condensate, the effective potential for the impurities $V_{\rm imp}(x) + \gamma |\Psi(x)|^2$ (blue dotted line), and the impurities for $V_0 = 30E_{\rm r}$ at $\gamma=1$. 
The condensate exhibits a more complicated density profile that is no longer a simple periodic structure.
In contrast, the density profile of the impurities closely resembles the profile for $V_0 = 0$ (as shown in panel (c2)) and remains unaffected despite the relatively deep depth of the underlying optical lattice.
As described by equation~\eqref{eq:SPSE}, the density of the impurities is determined by both the optical lattice and the BEC via the potential $V_{\rm imp}(x) + \gamma |\Psi(x)|^2$, and this is the one that now exhibits the periodic lattice structure commensurate to the number of TG bosons (blue dot-dashed line in panel (c1)).
Therefore, the impurities achieve the self-pinning state at $\gamma=1$, which leads to the opening of a band gap exclusively at $n = N$ in the corresponding single-particle energy spectrum that the self-pinning state admits (see panels (b1) and (b2)).

\section{\label{sec:quench_dynamics}Nonequilibrium properties of dimerized TG-impurities}

Equation~\eqref{eq:TFASE} represents a self-attractive $N$-CNLSE, in which 
a combination of effective mean-field attraction and the Pauli exclusion principle gives rise to bright soliton-train states for sufficiently large $\gamma$ when $V_{\rm imp} = V_{\rm BEC} = 0$~\cite{Gordon_1983, hiyane_2024}. In this section, we demonstrate that the ground state analyzed previously can be interpreted as dimerized soliton-trains.

\begin{figure}[tb]
    \centering
    \includegraphics[width=\linewidth]{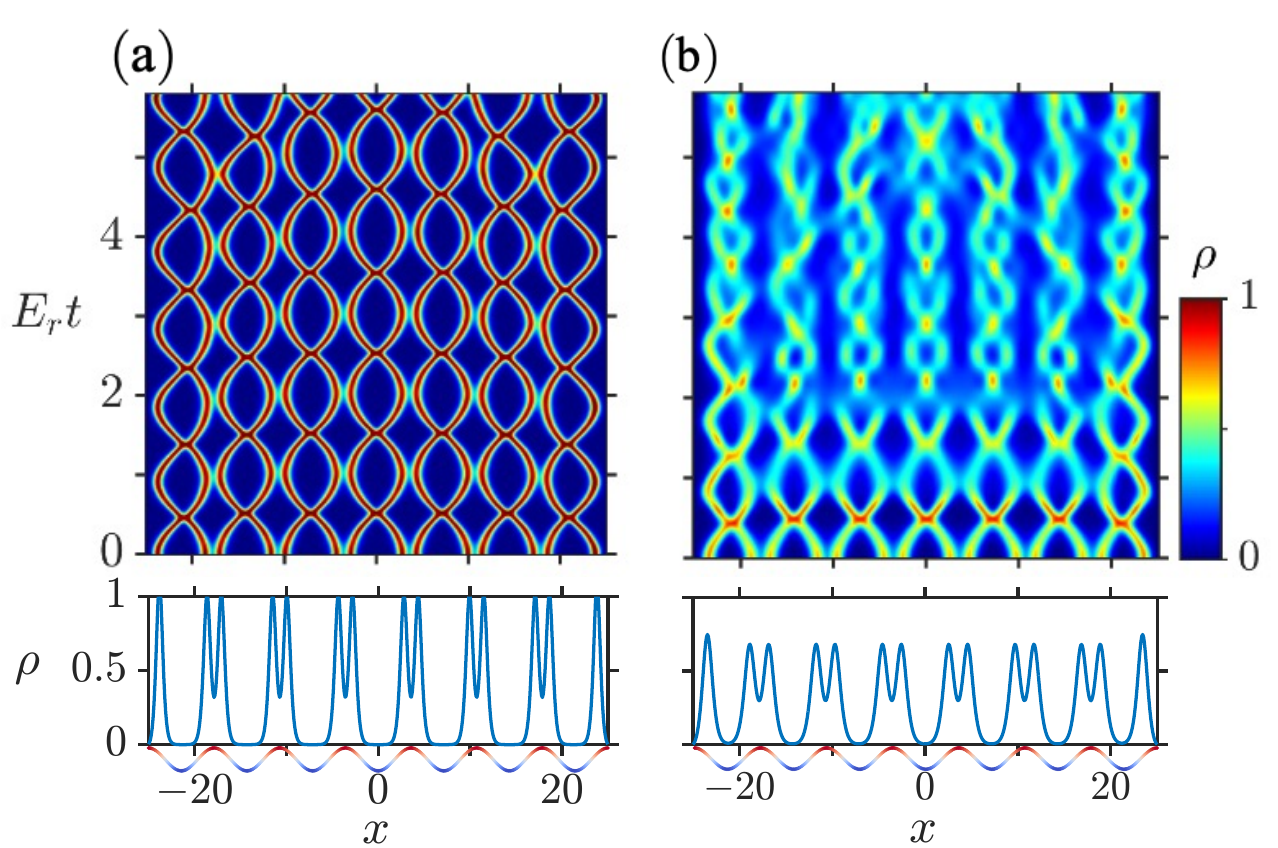}
    \caption{
     (Upper panel) time evolution of the density and (Lower panel) initial density of the impurities for (a) $\gamma=2$ and (b) $\gamma=1.6$, respectively.
     The dynamics is induced by removing the lattice trap only from the BEC at time $t=0$.
     The number of impurities is now $N=14$, and the number of lattice sites is $N_{\rm s}=7$ schematically depicted below the initial density profile.
     The lattice depth is $V_0=5E_{\rm r}$.
    }
    \label{fig:removefromBEC_density}
\end{figure}

To investigate this, we examine the quench dynamics of the impurities following the sudden removal of the optical lattice from the BEC, leaving it intact for the impurities. This specific quench protocol allows us to explore characteristic nonequilibrium properties across all values of $\gamma$. The boundary conditions remain the same as for the initial state: open boundaries for the impurities and closed boundaries for the BEC. We consider $N = 14$ impurities and $N_{\rm BEC} = 10^4$, with the optical lattice containing $N_{\rm s} = 7$ minima.

Three coupling regimes are of particular interest:
\begin{itemize}
    \item the weakly interacting regime $(\gamma \ll 1)$: the impurities are weakly coupled to the BEC bath as they move. Thus, removing the lattice potential from BEC does not significantly affect the impurities' dynamics.
    \item the crossover regime $(\gamma \sim 1)$: the dimers are not strongly bound, allowing overlap between neighboring dimers.
    \item the strongly coupled soliton regime $(\gamma \gg 1)$: the dimers are strongly bound and completely isolated from one another.
\end{itemize}

In the following, we first discuss the underlying nonequilibrium properties by studying the two representative coupling constants: $\gamma = 1.6$ 
and $\gamma = 2$.

In Fig.~\ref{fig:removefromBEC_density}(a), the density dynamics in the strongly coupled limit $(\gamma = 2)$ is presented.
Following the removal of the external optical lattice from the BEC, the effective single-particle potential for the impurities transitions from attractive to repulsive (see equation~\eqref{eq:effective_lattice_trap}).
As a result, each dimer splits into two isolated wave packets, which are accelerated by the repulsive optical lattice and which collide at each minimum of the external lattice potential, as observed in the density evolution in Fig.~\ref{fig:removefromBEC_density}(a).
At approximately $t_{\rm rev} \approx 0.7E_{\rm r}^{-1}$, each wave packet undergoes a quasi-revival into a dimer state. 
This revival time can be estimated using $t_{\rm rev} = \pi / \omega$, where $\omega \approx \sqrt{2V_0}k_{\rm r}$ is the effective harmonic oscillator frequency obtained from the first-order expansion of equation~\eqref{eq:Vopt}.
Throughout the time evolution, the density profile of the impurities remains localized due to the BEC-mediated self-attraction, which is the characteristic defining properties of the solitons~\cite{book_Lamb_1980}.
Although it is not shown here for brevity, the condensate’s density profile retains an overall Thomas–Fermi shape.

It is worth noting that the quasi-revival time is not identical for all localized solitonic impurities due to finite-size effects. For instance, the collision behavior of the two outermost impurities exhibits slight deviations. These small differences accumulate over longer time scales, and after a sufficiently long time $(t \gtrsim 4E_{\rm r}^{-1})$, the dynamics eventually deviates from simple harmonic motion.

In the crossover regime $(\gamma = 1.6)$, the nonequilibrium behavior differs significantly from the strong coupling limit.
The time evolution of the impurity density for $\gamma = 1.6$ is shown in Fig.~\ref{fig:removefromBEC_density}(b). One can see that initially, just after the removal of the optical lattice, the impurities remain localized and are accelerated by the external optical lattice, colliding at approximately half of the quasi-revival time, similar to the behavior observed for $\gamma = 2$. However, at the first quasi-revival, the impurities fail to fully recover their initially localized dimer states, appearing instead to become more delocalized.
This trend continues at the second quasi-revival, where the localized dimer state is destroyed, leading to significant delocalization over longer timescales.


\begin{figure}[tb]
    \centering
    \includegraphics[width=\linewidth]{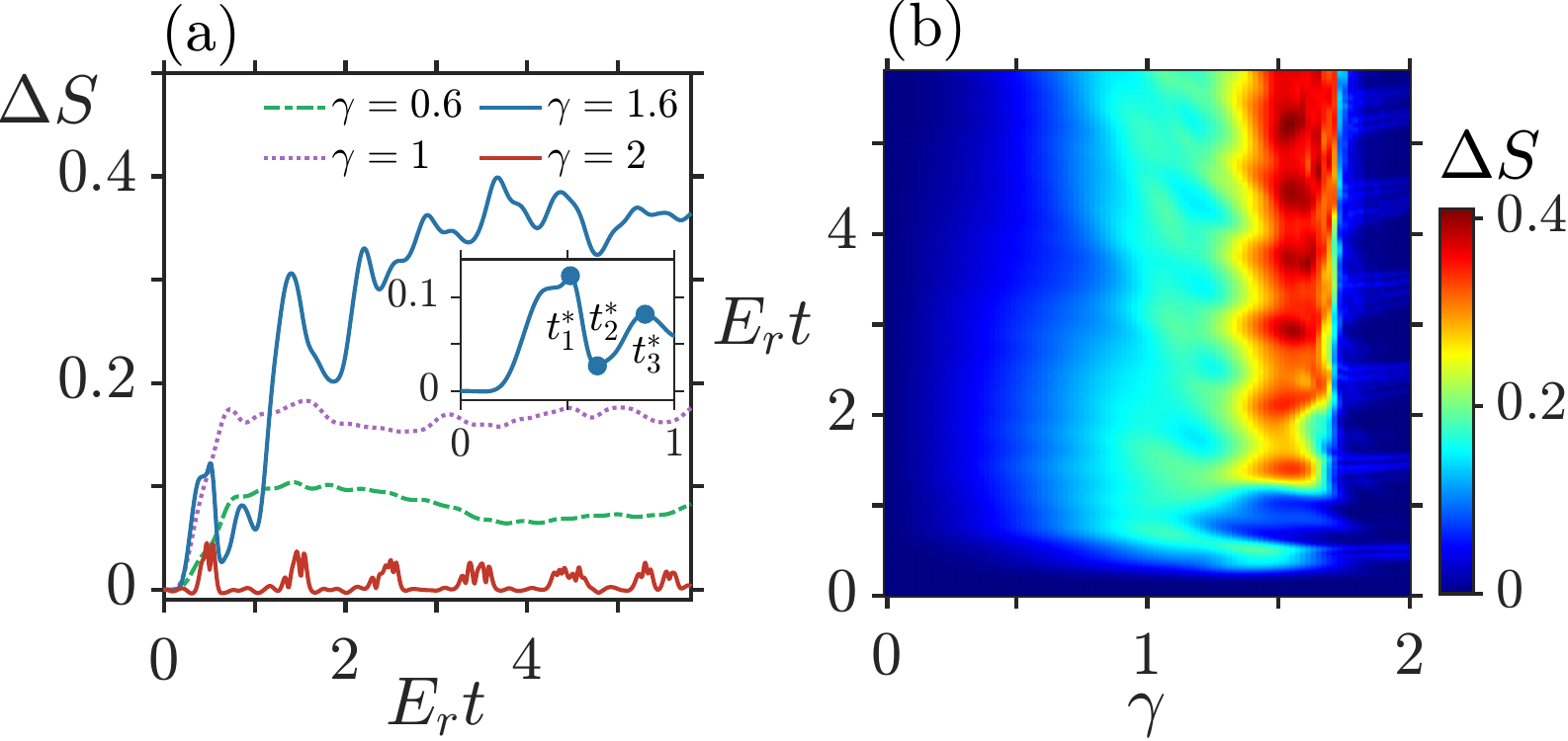}
    \caption{ Dynamics of the von-Neumann entropy $\Delta S(t)=S(t)-S(0)$ for (a) representative values for the coupling strengthes $\gamma$ and (b) as a function of $\gamma$ and $E_{\rm r}t$.
    Inset in (a) is the entropy within time $0<E_{\rm r}t<1$ for $\gamma=1.6$.
    }
    \label{fig:removefromBEC_entropy}
\end{figure}

To quantify this behavior, we compute the von Neumann entropy of the TG impurities, which characterises the correlations within the system. The von Neumann entropy is defined as $S(t)=-\sum_n\lambda_n(t)\ln \lambda_n(t)$ where  the $\lambda_n(t)$  are the occupation numbers of the reduced states  $\varphi_n(x,t)$, also known as natural orbitals~\cite{PezerBuljan,girardeauRSPDM}. 
These orbitals are defined as the eigenstates of the reduced single-particle density matrix (RSPDM)
\begin{equation}
    \int dx'\varrho(x,x',t)\varphi_n(x',t)=\lambda_n(t)\varphi_n(x,t),
\end{equation}
where the RSPDM is obtained by tracing out all particles except one~\cite{PezerBuljan,girardeauRSPDM}
\begin{equation}
    \varrho(x,x',t)=\int dx_2\cdots dx_N\Psi^*(x,x_2,\cdots, x_N,t)\Psi(x',x_2,\cdots, x_N,t)
\end{equation}
We show the entropy difference,  $\Delta S(t) = S(t) - S(t=0)$, as a function of time in Fig.~\ref{fig:removefromBEC_entropy}(a).

In the strong coupling regime ($\gamma = 2$), the entropy remains close to its initial value, except at  $t_{\rm rev}/2 \approx 0.35E^{-1}_{\rm r}$ when the impurities collide and the TG bosons start to become more mixed.
This behavior between collisions is expected, as strong localization throughout the evolution reduces wavefunction overlap, limiting the spread of correlations between the particles.
This leads to the characteristic soliton-train dynamics behaving as classical point-like particles observed in the density profile in Fig.~\ref{fig:removefromBEC_density} (a).
Then, for all cases except $\gamma = 2$, the entropy increases rapidly from its initial value, indicating that the reduced states of the impurities become more mixed.
This suggests that a larger number of natural orbitals than the initial state contribute to the dynamics, and the impurities are driven far from the initially localized state, leading to the delocalized density profiles observed in Fig.~\ref{fig:removefromBEC_density} (b).
After a sufficiently long time, the reduced state of the impurities relaxes to a highly mixed state due to the mean-field coupling to the BEC that induces the impurity-impurity interactions.

Let us now illustrate the full behavior of the entropy as a function of coupling strength and time, which is shown in Fig.~\ref{fig:removefromBEC_entropy}(b) serving as a dynamical phase diagram of the system.
The BEC and TG gas are nearly decoupled for the weak coupling limit $(\gamma \approx 0)$.
Thus, removing the optical lattice from the BEC has little to no effect on the impurities, resulting in $\Delta S(t) \approx 0$ throughout the dynamics.
As the coupling strength becomes stronger ($\gamma \sim 1)$, the entropy starts to increase over time and eventually saturates, indicating relaxation due to the interactions with the BEC bath.
However, increasing the coupling strength further, the system enters the nearly integrable soliton regime.
From the dynamical phase diagram, we identify that this transition occurs around $\gamma \sim 1.7$.
Then in this regime, the entropy remains nearly constant over time, reflecting the classical particle-like nature of the soliton-train.

It is worth commenting on the short-time behavior of the entropy around this transition point.
The inset of Fig.~\ref{fig:removefromBEC_entropy}(a) highlights this short-time behavior of $\Delta S$ for $\gamma = 1.6$.
One can see that initially, the entropy increases until it reaches a maximum at $t^*_1$, corresponding to the time when the collision of the impurities occurs. 
Then the entropy lowers down, and correspondingly, the impurities try to quasi-revive to the dimer state as observed in Fig.~\ref{fig:removefromBEC_density}(b).
However, unlike in the soliton regime, the entropy does not return to its initial value but instead reaches a finite minimum at $t^*_2$, after which it begins to rise again toward the quasi-revival time $t^*_3$. 
Then, the number of orbitals involved in the time evolution keeps increasing from the initial state, preventing the system from maintaining the quasi-integrable dynamics observed in Fig.~\ref{fig:removefromBEC_density}(a). Over time, the density profile alternates between localization and delocalization, manifesting as extrema in the entropy, and eventually becomes completely delocalized with accompanying the saturation of the entropy.

\section{\label{sec:conclusion}Conclusion}

In summary, the ground state and nonequilibrium properties of TG bosons in a BEC with an optical lattice are studied.
First, we examined the ground state properties of the impurity-BEC mixture in the optical lattice that selectively traps the impurities.
The impurities exhibit dimerization when the interspecies interaction is sufficiently strong due to the BEC-mediated attraction.
Then, we studied the impurity-BEC mixture, both confined in the optical lattice trap.
In the weak coupling limit ($\gamma \ll 1$), the effect of the lattice potential on the BEC is negligible in the arrangement of the TG dimers, and the impurities exhibit dimer state as in the species-selective lattice that only the impurities experience.
In the crossover regime ($\gamma \approx 1$), the BEC-mediated attractive lattice cancels the external repulsive optical lattice, resulting in a self-pinning state characterized by a single band gap at $n = N$. 
Increasing the interaction strength further, the effective lattice potential alters from repulsive ($\gamma < 1$) to BEC-mediated attractive ($\gamma > 1$), and the distribution of the dimers is rearranged accordingly.

When the optical lattice is suddenly removed from the BEC, the dimers break, and the impurities exhibit harmonic motion within the lattice wells. In the strong coupling limit ($\gamma = 2$), the impurities remain localized and display solitonic, quasi-integrable motion. However, in the intermediate coupling regime (finite but not strong coupling), the TG impurities thermalize within the BEC bath due to BEC-mediated self-interactions among the impurities.

The mean-field analysis carried out in this manuscript is useful in grasping the underlying physics with the simplest setup that neglects the correlation between TG bosons and the condensate.
However, it would be an interesting future work that extends the analysis presented here and sheds light on the role of the correlation between the two species by, e.g.\ the multi-configuration time-dependent Hartree method~\cite{Mistakidis2019Aug,Siegl2018May}.

\ack
This work was supported by the Okinawa Institute of Science and Technology Graduate University. The authors are grateful to the Scientific Computing and Data Analysis (SCDA) section of the Research Support Division at OIST for their invaluable assistance. 
This work was partially supported by the SIP Grant No. JPJ012367.
T.F. acknowledges support from JSPS KAKENHI Grant No. JP23K03290.
T.F. and T.B. are also supported by JST Grant No. JPMJPF2221.

\appendix

\section{Broken symmetry: Relation with the bosonic Josephson junction\label{appendix:double_well}}

\begin{figure}[tb]
    \centering
    \includegraphics[width=0.7\linewidth]{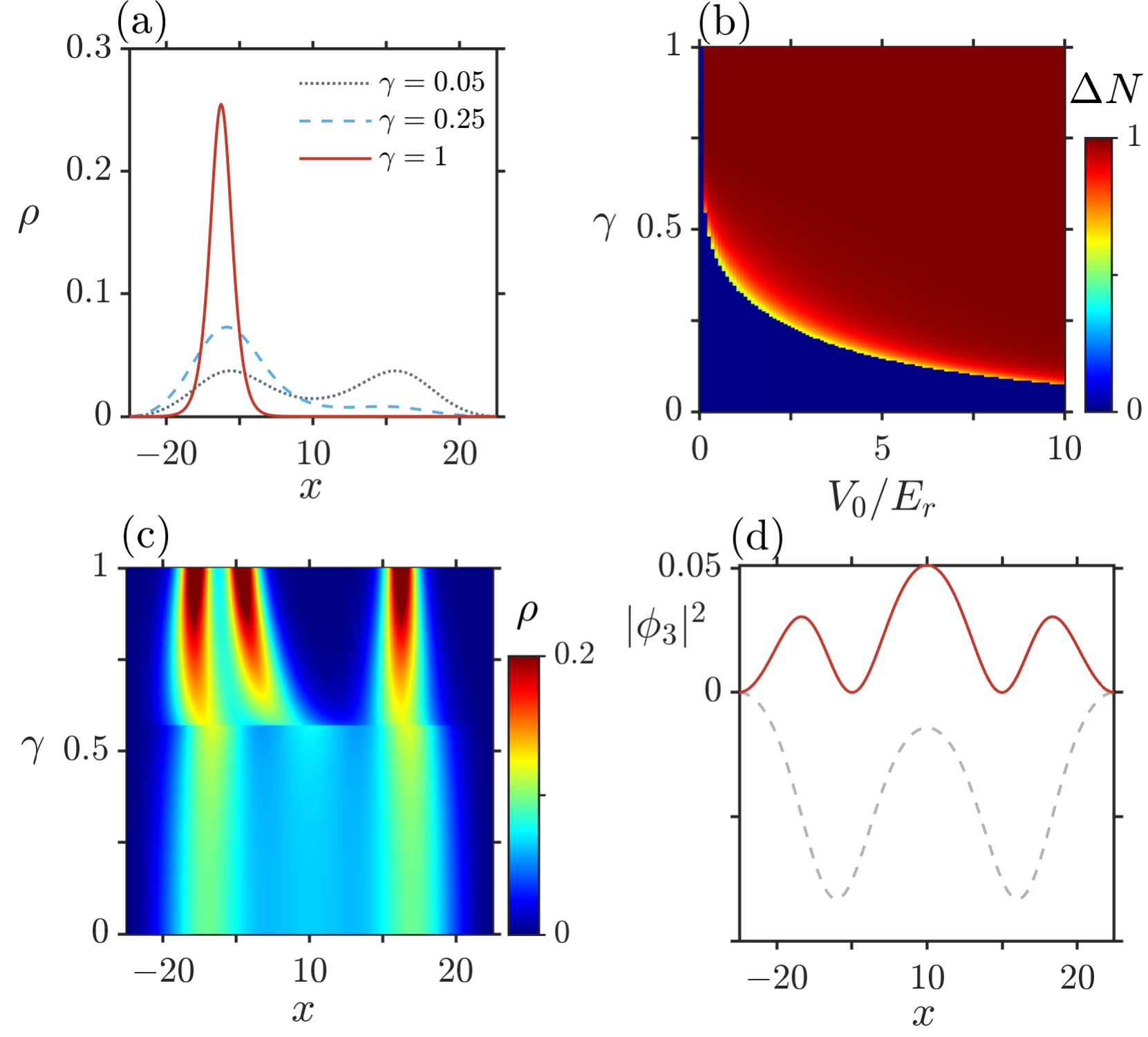}
    \caption{
    Impurities trapped in a double-well potential immersed in BEC.
    (a) Density of the single impurity for the representative inter-species interaction and (b) corresponding occupation imbalance $\Delta N=|n_{\rm L}-n_{\rm R}|$, where $n_{\rm L(R)}$ is the occupation probabilities at site $x<0 (x>0)$.
    (c) Density of the three impurities in a function of $x$ and $\gamma$, and (d) the highest energy state $|\phi_3|^2$.
    The grey dashed line in (c) is the effective potential for $\phi_3$: $\mathcal{V}=V_{\rm opt}-g_{\rm eff}(|\phi_1|^2+|\phi_2|^2)$.
    The lattice depth for the panel (a), (c), and (d) is $V_0=2.5E_r$.
    } 
    \label{fig:SSB_N1N3}
\end{figure}

In Fig.~\ref{fig:speciesselective_density}(h), we have shown that emerging dimer states can break the left-right symmetry for the $N=9$ impurities species-selectively trapped in $N_{\rm s}=5$ lattice sites systems.
In this Appendix, we discuss this spontaneous symmetry breaking by considering the minimum model of  $N_{\rm s}=2$ (double-well trap) with impurity numbers of $N=1$ and $N=3$, along with the relation between impurity-BEC mixture and the bosonic Josephson junction.

Let us first discuss a single impurity in the species-selective double-well trap immersed in the BEC.
When there is no coupling to the BEC $\gamma=0$, the lowest energy solution of the system is symmetric $(\psi_L(x)+\psi_R(x))/\sqrt{2}$, where $\psi_{L(R)}(x)$ is the localized state at left (right) side of the well.
Fig.~\ref{fig:SSB_N1N3}(a) shows the probability densities of an impurity in the BEC for the three representative values of the interaction strength.
As can be seen, for weak coupling, $\gamma=0.05$, the above symmetric solution still holds.
However, perhaps unexpectedly, the impurity tends to occupy the left lattice site as the coupling strength increases, resulting in a symmetry broken ground state.

This spontaneous symmetry breaking can be understood from the effective Thomas--Fermi description of the NLSE in equation~\eqref{eq:TFASE}.
For $N=1$, the effective NLSE has the the form of an attractively self-interacting system in the double well trap, which is exactly the same governing equation as for an attractive condensate in the double well potential.
Such systems have been extensively investigated in the context of the bosonic Josephson junction~\cite{Smerzi_1997,Raghavan_1999,Mazzarella_2010,wysocki_2024}, and the symmetric ground state solution is known to be unstable for sufficiently strong attraction.
This is driven by the self-attraction being so strong that the localized solution is energetically favorable.
Such localized solution of the condensate (in our case, impurity) either at the site of $x<0$ or $x>0$ possesses degenerate energy, leading to the spontaneous localization to the left or right lattice site~\cite{Smerzi_1997,Raghavan_1999,Mazzarella_2010}.
We note that in our calculations the spontaneous symmetry breaking is controlled to localize the atom at a site $x<0$ by applying a small linear potential $V_L=10^{-10}x$, which is orders of magnitudes smaller than any other degrees of freedom in the system.

For reference, we show the density imbalance $\Delta N=|\int^L_0dx\rho-\int^0_{-L}dx\rho|$ as a function of $\gamma$ and $V_0$ in Fig.~\ref{fig:SSB_N1N3}(b).
At $V_0=0$, there is no symmetry breaking solution as expected and $\Delta N=0$ throughout $\gamma$.
For finite $V_0$ and weak coupling, the probability density of the impurity is symmetric and $\Delta N=0$.
The symmetry breaking occurs earlier for larger $V_0$ since the impurity is localized more which enhances the self-interaction.

Based on this observation, we also discuss the case for $N=3$ and $N_{\rm s}=2$ with the symmetry-breaking linear potential introduced above.
The density of the TG gas is presented in Fig.~\ref{fig:SSB_N1N3}(c) for $V_0=2.5E_{\rm r}$, which shows the emergence of the doubly occupied dimer state by breaking the left-right symmetry spontaneously.
The mechanism of this symmetry-breaking can again be explained through the mediated self-attraction among the impurities.
For instance, the governing equation for the $\phi_3$ can be written in terms of the NLSE by defining the effective potential: $\mathcal{V}=-\gamma^2(|\phi_1|^2+|\phi_2|^2)+V_{\rm opt}+V_{\rm L}$.
Then, $N$-CNLSE~\eqref{eq:TFASE} reduces to the NLSE for $\phi_3$ as 
\begin{align}
    E_3\phi_3(x)=
    \biggl(-\frac{1}{2}\pdv[2]{x}+\mathcal{V}(x)-\gamma^2|\phi_3(x)|^2\biggr)\phi_3(x).
\end{align}
We show $\mathcal{V}$ in Fig.~\ref{fig:SSB_N1N3}(d) as the dashed line, which clearly forms the effective double-well potential for states $\phi_3$, and the state $\phi_3$, shown as a solid line, is symmetric.
As we exemplified for the case of $N=1$, such a symmetric solution is unstable when the BEC-mediated self-attraction $-\gamma|\phi_3|^2$ is sufficiently strong, and spontaneous localization in one of the (effective) lattice sites occurs (controlled to be the site at $x<0$ by $V_{\rm L}$), keeping the node at the lattice minimum due to the Pauli exclusion principle.
Although we do not present it here, the density imbalance in the function of $\gamma $ and $V_0$ follows a similar trend as Fig.~\ref{fig:SSB_N1N3}(b).
In both cases, for any small but finite $V_0$, the symmetry-breaking solution exists for sufficiently strong inter-species coupling $\gamma$, indicating the instability due to the self-localization.

So far, we have considered a minimum model of lattice potential $N_{\rm s}=2$. However, the above discussion can be readily applied to the case of $N_{\rm s}>2$.
The symmetric ground state is unstable for any $N_{\rm s}\geq2$ and $V_0>0$, when the BEC-mediated attraction is sufficiently strong, and impurities prefer to localize and form dimer states by spontaneously breaking the discrete translational symmetry.
This symmetry breaking is controlled by the box trap in case of the setup Fig.~\ref{fig:speciesselective_density}(a)-(g,i) and by the vanishingly small linear ramp in case of the setup Fig.~\ref{fig:speciesselective_density}(h).

Finally, let us comment on the detailed profile of the single particle states presented in  Fig.~\ref{fig:speciesselective_density}(d).
As shown in Fig.~\ref{fig:speciesselective_density} (d), the presence of the dimer state at the center of the box causes the $n = 2$ to $n = 5$ states to split into two segments, located in the left (green shaded domain in $x < 0$) and right (yellow shaded domain in $x > 0$) regions.
Due to the strong self-attraction among impurities, each of the four impurities prefers to localize at a single site, filling the four outer lattice sites nearly individually.

\section*{References}
\bibliographystyle{iopart-num}

\bibliography{Manuscript}

\providecommand{\newblock}{}
\begin{thebibliography}{10}
\expandafter\ifx\csname url\endcsname\relax
  \def\url#1{{\tt #1}}\fi
\expandafter\ifx\csname urlprefix\endcsname\relax\def\urlprefix{URL }\fi
\providecommand{\eprint}[2][]{\url{#2}}

\bibitem{Giamarchi_2003}
Giamarchi T 2003 {\em {Quantum Physics in One Dimension}\/} (Oxford University Press) ISBN 9780198525004

\bibitem{pitaevskii_stringari2016}
Pitaevskii L and Stringari S 2016 {\em Bose-Einstein condensation and superfluidity\/} vol 164 (Oxford University Press)

\bibitem{Mistakidis2023Nov}
Mistakidis S~I, Volosniev A~G, Barfknecht R~E, Fogarty T, Busch {\relax Th}, Foerster A, Schmelcher P and Zinner N~T 2023 {\em Phys. Rep.\/} {\bf 1042} 1--108 ISSN 0370-1573

\bibitem{Girardeau:60}
Girardeau M 1960 {\em Journal of Mathematical Physics\/} {\bf 1} 516--523

\bibitem{Buchler_2003}
Büchler H~P, Blatter G and Zwerger W 2003 {\em Phys. Rev. Lett.\/} {\bf 90}(13) 130401

\bibitem{haller_2010}
Haller E, Hart R, Mark M~J, Danzl J~G, Reichs{\"o}llner L, Gustavsson M, Dalmonte M, Pupillo G and N{\"a}gerl H~C 2010 {\em Nature\/} {\bf 466} 597--600

\bibitem{Book:Popov_1983}
Popov V~N 1983 {\em {Functional Integrals in Quantum Field Theory and Statistical Physics}\/} (Dordrecht, The Netherlands: Springer Netherlands) ISBN 978-90-277-1471-8

\bibitem{Mora_2003}
Mora C and Castin Y 2003 {\em Phys. Rev. A\/} {\bf 67}(5) 053615

\bibitem{McGuire1964May}
McGuire J~B 1964 {\em J. Math. Phys.\/} {\bf 5} 622--636 ISSN 0022-2488

\bibitem{Carr_2000_2}
Carr L~D, Clark C~W and Reinhardt W~P 2000 {\em Phys. Rev. A\/} {\bf 62}(6) 063611

\bibitem{Kanamoto_2005}
Kanamoto R, Saito H and Ueda M 2005 {\em Phys. Rev. Lett.\/} {\bf 94}(9) 090404

\bibitem{Smerzi_1997}
Smerzi A, Fantoni S, Giovanazzi S and Shenoy S~R 1997 {\em Phys. Rev. Lett.\/} {\bf 79}(25) 4950--4953

\bibitem{Raghavan_1999}
Raghavan S, Smerzi A, Fantoni S and Shenoy S~R 1999 {\em Phys. Rev. A\/} {\bf 59}(1) 620--633

\bibitem{Mazzarella_2010}
Mazzarella G and Salasnich L 2010 {\em Phys. Rev. A\/} {\bf 82}(3) 033611

\bibitem{wysocki_2024}
Wysocki P, Jachymski K, Astrakharchik G~E and Tylutki M 2024 {\em Phys. Rev. A\/} {\bf 110}(3) 033303

\bibitem{Josephson1962Jul}
Josephson B~D 1962 {\em Physics Letters\/} {\bf 1} 251--253 ISSN 0031-9163

\bibitem{Book:Mahan2000}
Mahan G~D 2000 {\em {Many-Particle Physics}\/} (Springer US) ISBN 978-1-4757-5714-9

\bibitem{Karpiuk_2004}
Karpiuk T, Brewczyk M, Ospelkaus-Schwarzer S, Bongs K, Gajda M and Rza\ifmmode \mbox{\c{}}\else \c{}\fi{}\ifmmode~\dot{z}\else \.{z}\fi{}ewski K 2004 {\em Phys. Rev. Lett.\/} {\bf 93}(10) 100401

\bibitem{Salerno_2005}
Salerno M 2005 {\em Phys. Rev. A\/} {\bf 72}(6) 063602

\bibitem{Karpiuk_2006}
Karpiuk T, Brewczyk M and Rza\ifmmode \mbox{\c{}}\else \c{}\fi{}\ifmmode~\dot{z}\else \.{z}\fi{}ewski K 2006 {\em Phys. Rev. A\/} {\bf 73}(5) 053602

\bibitem{Santhanam_2006}
Santhanam J, Kenkre V~M and Konotop V~V 2006 {\em Phys. Rev. A\/} {\bf 73}(1) 013612

\bibitem{DeSalvo_2017}
DeSalvo B~J, Patel K, Johansen J and Chin C 2017 {\em Phys. Rev. Lett.\/} {\bf 119}(23) 233401

\bibitem{Rakshit_2019_NJP}
Rakshit D, Karpiuk T, Zin P, Brewczyk M, Lewenstein M and Gajda M 2019 {\em New Journal of Physics\/} {\bf 21} 073027

\bibitem{Rakshit_2019_scipost}
Rakshit D, Karpiuk T, Brewczyk M and Gajda M 2019 {\em SciPost Phys.\/} {\bf 6} 079

\bibitem{Desalvo_2019}
DeSalvo B~J, Patel K, Cai G and Chin C 2019 {\em Nature\/} {\bf 568} 61--64

\bibitem{Keller_2022}
Keller T, Fogarty T and Busch T 2022 {\em Phys. Rev. Lett.\/} {\bf 128}(5) 053401

\bibitem{hiyane_2024}
Hiyane H, Busch T and Fogarty T 2024 {\em Phys. Rev. Res.\/} {\bf 6}(3) L032040

\bibitem{Zschetzsche_2024}
Zschetzsche L and Zillich R~E 2024 {\em Phys. Rev. Res.\/} {\bf 6}(2) 023137

\bibitem{Mikkelsen_2018}
Mikkelsen M, Fogarty T and Busch T 2018 {\em New Journal of Physics\/} {\bf 20} 113011

\bibitem{Gordon_1983}
Gordon J~P 1983 {\em Opt. Lett.\/} {\bf 8} 596--598

\bibitem{book_Lamb_1980}
Lamb G~L 1980 {\em Elements of soliton theory\/} (Wiley New York) ISBN 0471045594

\bibitem{PezerBuljan}
Pezer R and Buljan H 2007 {\em Phys. Rev. Lett.\/} {\bf 98}(24) 240403

\bibitem{girardeauRSPDM}
Girardeau M~D, Wright E~M and Triscari J~M 2001 {\em Phys. Rev. A\/} {\bf 63}(3) 033601

\bibitem{Mistakidis2019Aug}
Mistakidis S~I, Hilbig L and Schmelcher P 2019 {\em Phys. Rev. A\/} {\bf 100} 023620

\bibitem{Siegl2018May}
Siegl P, Mistakidis S~I and Schmelcher P 2018 {\em Phys. Rev. A\/} {\bf 97} 053626

\end{thebibliography}

\end{document}